\documentclass{elsart}

\usepackage{graphicx}
\usepackage{ifthen}
\usepackage{xspace}

\usepackage{amsmath}

\begin{document}

\begin{frontmatter}
\title{Study of Hadronic Five-Body Decays of Charmed Mesons}

\date{\today}

The FOCUS Collaboration

\author[ucd]{J.~M.~Link}
\author[ucd]{M.~Reyes}
\author[ucd]{P.~M.~Yager}
\author[cbpf]{J.~C.~Anjos}
\author[cbpf]{I.~Bediaga}
\author[cbpf]{C.~G\"obel}
\author[cbpf]{J.~Magnin}
\author[cbpf]{A.~Massafferri}
\author[cbpf]{J.~M.~de~Miranda}
\author[cbpf]{I.~M.~Pepe}
% ...

\address[ucd]{University of California, Davis, CA 95616} 
\address[cbpf]{Centro Brasileiro de Pesquisas F\'\i sicas, Rio de Janeiro, RJ, Brasil} 
\address[cinv]{CINVESTAV, 07000 M\'exico City, DF, Mexico} 
\address[cu]{University of Colorado, Boulder, CO 80309} 
\address[fnal]{Fermi National Accelerator Laboratory, Batavia, IL 60510} 
\address[fras]{Laboratori Nazionali di Frascati dell'INFN, Frascati, Italy I-00044} 
\address[ui]{University of Illinois, Urbana-Champaign, IL 61801} 
\address[iu]{Indiana University, Bloomington, IN 47405} 
\address[korea]{Korea University, Seoul, Korea 136-701} 
\address[korea2]{Kyungpook National University, Taegu, Korea 702-701}
\address[milan]{INFN and University of Milano, Milano, Italy} 
\address[nc]{University of North Carolina, Asheville, NC 28804} 
\address[pavia]{Dipartimento di Fisica Nucleare e Teorica and INFN, Pavia, Italy} 
\address[pr]{University of Puerto Rico, Mayaguez, PR 00681} 
\address[sc]{University of South Carolina, Columbia, SC 29208} 
\address[ut]{University of Tennessee, Knoxville, TN 37996} 
\address[vu]{Vanderbilt University, Nashville, TN 37235} 
\address[wisc]{University of Wisconsin, Madison, WI 53706}

\endnote{\small See http://www-focus.fnal.gov/authors.html for
additional author information}

 
\begin{abstract}
We study
the decay of $D^+$ and $D^+_s$ mesons into charged five-body final states, and
report the discovery of the decay mode $D^+\rightarrow
K^+K^-\pi^+\pi^+\pi^-$, as well as measurements of the branching ratios of the
decay modes 
$D^+\rightarrow K^-\pi^+\pi^+\pi^+\pi^-$,
$D_s^+\rightarrow K^+K^-\pi^+\pi^+\pi^-$,
$D_s^+\rightarrow \phi\pi^+\pi^+\pi^-$ and
$D^+$/$D_s^+\rightarrow \pi^+\pi^+\pi^+\pi^-\pi^-$. An analysis of the 
resonant substructure for $D^+ \rightarrow K^-\pi^+\pi^+\pi^+\pi^-$ and 
$D_s^+\rightarrow K^+K^-\pi^+\pi^+\pi^-$ 
is also included, with evidence suggesting that both decays proceed primarily
through an $a_1$ vector resonance.  
  
\end{abstract}
\end{frontmatter}



The hadronic five-body decays of charmed mesons have been studied in
recent years~[1--6], but limited statistics have prevented
precise measurements of their resonant substructure. Theoretical predictions 
are limited mainly to two-body
decay modes, and little is known about how five-body final states are produced.
Theoretical discussion suggests a ``vector-dominance model,''
in which heavy flavor mesons decay into a two-body intermediate state 
by emitting a $W$, which
immediately hadronizes into a charged vector, axial vector, or pseudoscalar
meson~[7]. 
The charged meson then decays strongly to produce a many-body 
final state. Confirmation of this model could provide a mechanism for the 
production of five-body final states.
 
The FOCUS Collaboration~[8--10] has studied 
two five-body decay modes, 
$D^+\!\rightarrow K^-\pi^+\pi^+\pi^+\pi^-$ and 
$D_s^+\!\rightarrow K^+K^-\pi^+\pi^+\pi^-$. We find evidence that 
in both modes the resonant substructure is dominated by a two-body vector 
resonance involving the $a_1(1260)^+$. We also present
inclusive branching ratio 
measurements of four charged five-body hadronic decays, 
including the first evidence of the decay mode 
$D^+\!\rightarrow K^+K^-\pi^+\pi^+\pi^-$.

Five-body $D^+$ and $D_s^+$ decays are reconstructed using a candidate driven
vertex algorithm~[8]. A decay vertex is formed from the five reconstructed
tracks. The momentum vector of the parent $D$ meson is then used as a seed to
intersect other tracks in order to find the production vertex. 
Events 
are selected based on a number of criteria.
The confidence level of the decay vertex is required to be greater than 
1$\%$. The confidence level that a track from the decay vertex 
intersects the production vertex is required to be less than 1$\%$. 
The likelihood for each particle to be a proton,
kaon, pion, or electron based on \v{C}erenkov particle identification is used to make
additional requirements~[9]. For each kaon candidate we
require the negative log-likelihood kaon hypothesis, 
$W_K = -2 \ln (\textrm{kaon likelihood})$, 
to be favored over the corresponding pion hypothesis $W_\pi$ by 
$W_\pi - W_K > 3$. In addition, for each pion candidate 
we require the pion hypothesis to be favored
over any alternative hypothesis. We also require the significance of separation
of the production and decay
vertices to be at least 10. 
In order to reduce background
due to secondary interactions of particles from the 
 production vertex, we require the $D$ reconstructed momentum to be
greater than 25 GeV/$c$ and the secondary vertex to be outside of the
target. Finally, we remove events that are consistent with various
$D^{*}$ decays.

We turn now to additional analysis cuts made in individual modes, beginning
with  $D^+ \!\rightarrow K^-\pi^+\pi^+\pi^+\pi^-$. Because this mode is the most
abundant we apply only the standard cuts used in all modes.
Figure~1a shows the $K4\pi$ invariant mass plot. The distribution is fitted with a Gaussian 
for the $D^+$
signal  and a $2^{\textrm{nd}}$ degree polynomial 
for the background. A binned maximum likelihood fit gives $2923\pm78$ events.

The $D^+$/$D_s^+ \!\rightarrow \pi^+\pi^+\pi^+\pi^-\pi^-$ modes are more difficult
to detect, due to the large combinatorial background. To reduce this
background we increase the separation of the secondary vertex from the
target to two standard deviations.  We further impose a series of selection cuts to 
remove misidentified charm decays. We remove the decays
$D^+$/$D_s^+\!\rightarrow\eta^{\prime}\pi^+$, $\eta^{\prime}\!\rightarrow
\pi^+\pi^+\pi^-\pi^-\pi^0$ by requiring the four pion reconstructed mass to be
larger than the $\eta^{\prime}-\pi^0$ mass difference, that is, 
$M_{4\pi}>0.825$ GeV/$c^2$. Figure~1b shows the five-pion invariant mass plot
for events that satisfy these
cuts. The distribution is fitted with a Gaussian for the $D^+$ signal
($835\pm49$ events), another Gaussian for the $D_s^+$ signal ($671\pm47$ events) 
and a $1^{\textrm{st}}$ degree polynomial for the 
background. 

For the $D_s^+\!\rightarrow K^+K^-\pi^+\pi^+\pi^-$ mode the requirement of two
kaons in the final state greatly reduces background, 
allowing us to apply only the standard cuts used in all
modes. Figure~1c shows the 
$K^+K^-\pi^+\pi^+\pi^-$ invariant mass plot for events satisfying these cuts. 
We fit to a Gaussian ($240\pm30$ events) and $2^{\textrm{nd}}$ degree polynomial.

For the $K^+K^-\pi^+\pi^+\pi^-$ final state we have also studied the subresonant
decay $D_s^+\!\rightarrow \phi\pi^+\pi^+\pi^-$, by additionally requiring the
$K^+K^-$ invariant mass combination to be consistent with the $\phi$
mass.
The $\phi\pi^+\pi^+\pi^-$ invariant mass plot is shown in Figure~1e. We
fit to a Gaussian ($136\pm14$ events) and $2^{\textrm{nd}}$ degree polynomial.

The decay $D^+\!\rightarrow K^+K^-\pi^+\pi^+\pi^-$ is Cabibbo suppressed. 
We require a significance of vertex separation of 20,
and tighten particle
identification cuts on both kaons to $W_\pi-W_K > 4$, but remove all 
requirements on the pions. We also require the $D^+$ reconstructed momentum to
be greater than 50 GeV/$c$. Figure~1d shows the resulting $K^+K^-\pi^+\pi^+\pi^-$ 
invariant mass plot. This is the first observation of this mode.
We fit with a Gaussian for the $D^+$ signal ($38\pm8$ events), 
another Gaussian for the $D_s$ events and a $2^{\textrm{nd}}$ degree polynomial 
for the background.

We measure the branching fraction of the $D^+\!\rightarrow
K^-\pi^+\pi^+\pi^+\pi^-$ mode relative to $D^+\!\rightarrow 
K^-\pi^+\pi^+$, then measure the branching fractions of the other $D^+$ modes 
relative to the $D^+\!\rightarrow K^-\pi^+\pi^+\pi^+\pi^-$ to reduce 
systematic effects due to differences in the number of decay products.
All $D_s^+$ decay modes are measured relative to $D_s^+\!\rightarrow K^+K^-\pi^+$.
The normalizing decay modes are subjected to the same vertex cuts and analogous
\v{C}erenkov identification cuts as the mode in question to minimize
systematic errors. The detector and analysis efficiency is calculated using a
Monte Carlo simulation. For modes included in our resonant substructure
analysis the Monte Carlo contains the incoherent mixture of subresonant 
decays determined
by our analysis. For modes not included in our resonant substructure analysis, 
the Monte Carlo is
composed of five-body phase space. 
We test for dependency on cut selection by individually
varying each cut. The
results, compared with existing measurements, are shown in Table~1. 
\begin{figure}[t]
\begin{center}
\includegraphics[width=4.5in,clip=true]{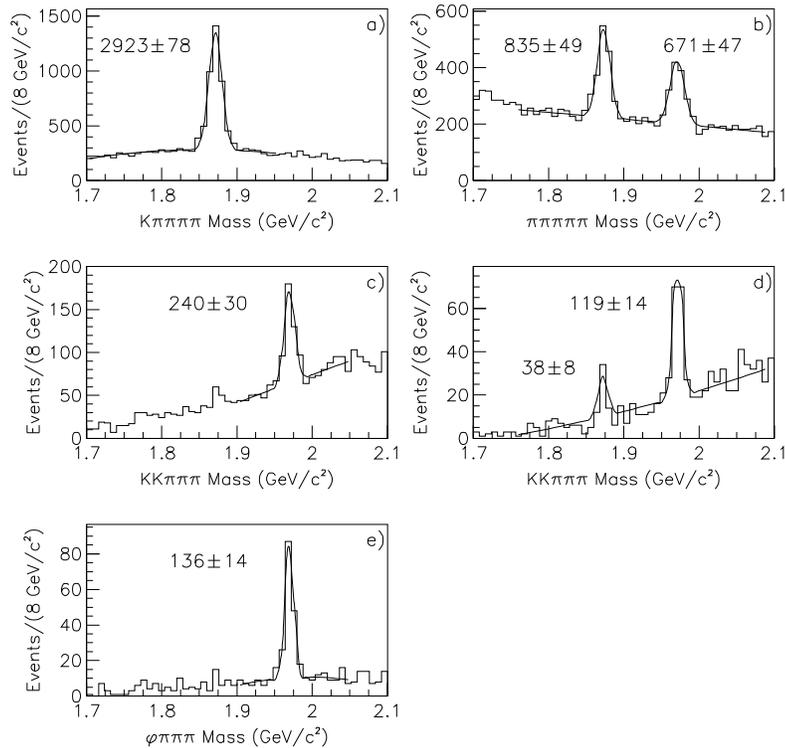}
\caption{(a) $K4\pi$ invariant mass distribution. (b) 5$\pi$ 
invariant mass distribution. (c) $KK3\pi$ invariant mass distribution for
$D_s^+$ optimized cuts.
(d) $KK3\pi$ invariant mass distribution for $D^+$ optimized cuts. (e) $\phi 3\pi$ 
invariant mass distribution.The fits are described in the text and the numbers
quoted are the yields.}
\label{Fig 1}
\end{center}
\end{figure}

We studied systematic effects due to uncertainties in Monte Carlo simulation,
fitting procedure, resonant substructure, Monte Carlo statistics and absolute
tracking efficiency. 
To determine the systematic error we follow a procedure based on the S-factor
method used by the Particle Data Group [11]. For each mode we split the data 
sample into four independent
subsamples based on $D$ momentum and period of time in which the data was
collected.  We then define the split sample variance as the difference between
the scaled variance and the statistical variance if the former exceeds the
latter. We also evaluate systematic effects associated
with Monte Carlo simulation of multi-body decays. The branching ratios are
evaluated with multiple conditions on the isolation of the production vertex, with
the variance used as the systematic error. 
In addition we evaluate the systematic effects based on different
fitting procedures. The branching ratios are evaluated under various fit
conditions, and the variance is used as the systematic error, as all fit
variants are a priori equally likely. We also evaluate systematic effects
due to uncertainty in resonant substructure by calculating the branching
ratios using various mixtures of subresonant states in the Monte Carlo. The
variance in the branching ratios for different subresonant mixtures is used as the
systematic error, treating each subresonant mixture as a priori equally likely. We also 
evaluate the systematic effect from
Monte Carlo statistics, adding in quadrature the uncertainty in the calculated 
efficiencies from both the signal and normalizing mode. Finally we
evaluate systematic effects from uncertainty in absolute tracking efficiency 
of multi-body decays using studies of $D^0\!\rightarrow K^-\pi^+\pi^+\pi^-$ and
$D^0\!\rightarrow K^-\pi^+$ decays. The systematic effects are then all added 
together in quadrature to obtain the final systematic error.
\begin{table}[t]   
\begin{center}
\caption{Branching ratios for five-body modes and comparison to the previous
measurements by E687. All branching ratios are inclusive of subresonant modes.}
\begin{tabular}{cccc} \hline \hline
Decay Mode&FOCUS&E687[6]\\ 
\hline 
$\frac{\Gamma(D^+\!\rightarrow K^-\pi^+\pi^+\pi^+\pi^-)}{\Gamma(D^+\!\rightarrow
K^-\pi^+\pi^+)}$&0.058$\pm$0.002$\pm$0.006&0.077$\pm$0.008$\pm$0.010\\
$\frac{\Gamma(D^+\!\rightarrow \pi^+\pi^+\pi^+\pi^-\pi^-)}{\Gamma(D^+\!\rightarrow
K^-\pi^+\pi^+\pi^+\pi^-)}$&0.290$\pm$0.017$\pm$0.011&0.299$\pm$0.061$\pm$0.026\\
$\frac{\Gamma(D_s^+\!\rightarrow \pi^+\pi^+\pi^+\pi^-\pi^-)}{\Gamma(D_s^+\!\rightarrow
K^-K+\pi^+)}$&0.145$\pm$0.011$\pm$0.010&0.158$\pm$0.042$\pm$0.031\\
$\frac{\Gamma(D_s^+\!\rightarrow K^+K^-\pi^+\pi^+\pi^-)}{\Gamma(D_s^+\!\rightarrow
K^-K^+\pi^+)}$&0.150$\pm$0.019$\pm$0.025&0.188$\pm$0.036$\pm$0.040\\
$\frac{\Gamma(D_s^+\!\rightarrow \phi\pi^+\pi^+\pi^-)}{\Gamma(D_s^+\!\rightarrow
\phi\pi^+)}$&0.249$\pm$0.024$\pm$0.021&0.28$\pm$0.06$\pm$0.01\\
$\frac{\Gamma(D^+\!\rightarrow K^+K^-\pi^+\pi^+\pi^-)}{\Gamma(D^+\!\rightarrow
K^-\pi^+\pi^+\pi^+\pi^-)}$&0.040$\pm$0.009$\pm$0.019&\\
\hline
\hline
\end{tabular}
\end{center}
\end{table}

In addition to reporting inclusive branching ratio measurements,
we have studied the resonance substructure in two decays:
$D^+ \!\rightarrow K^-\pi^+\pi^+\pi^+\pi^-$ and $D_s^+\!\rightarrow K^+K^-\pi^+\pi^+\pi^-$. 
We use an incoherent binned fit method, also used in reference 6,
 a simplified approach which assumes the
final state is an incoherent superposition of subresonant decay modes containing
vector resonances. A coherent analysis for decays into five-body final states
 has not yet been 
attempted, and would be very difficult given the statistics of this experiment. 
 For the $D^+ \!\rightarrow K^-\pi^+\pi^+\pi^+\pi^-$ mode we consider the 
lowest mass ($K^-\pi^+$) and
($\pi^+\pi^-$) resonances, as well as a nonresonant channel: 
$\overline{K^{*0}}\pi^-\pi^+\pi^+$, $K^-\rho^0\pi^+\pi^+$,    
$\overline{K^{*0}}\rho^0\pi^+$, and $(K^-\pi^+\pi^+\pi^+\pi^-)_{\textrm{NR}}$. All 
states not explicitly considered are assumed to be included in the 
nonresonant channel.

We determine the acceptance corrected yield into each subresonant mode using a
weighting technique whereby each event is weighted by its kinematic values
in three submasses: 
($K^-\pi^+$), ($\pi^+\pi^-$), and ($\pi^+\pi^+$). No resonance in the 
($\pi^+\pi^+$) submass exists, but we include it in order to compute a
meaningful $\chi^2$ estimate of the fit. The weights are obtained
using separate Monte Carlo simulations for the four decay modes, with the
Particle Data Group values~[11] for the mass
and width of each resonance. Eight
kinematic bins are constructed depending on whether each of the three submasses 
falls within the expected resonance (In the case of $\pi^+\pi^+$, the
bin is split into high and low mass regions). For each Monte Carlo simulation
the bin population in the eight bins is determined using a sideband
subtracted cut on the $D^+$ peak, allowing a linear
transformation matrix to be calculated. The weights are then determined 
from the transformation matrix by a $\chi^2$ minimization procedure. Each data
event which satisfies our selection cuts is then weighted according to its
kinematic values in the submass bins. 
Once the weighted distributions for each of the
four modes are generated, we determine the acceptance corrected yield by fitting
the distributions with a Gaussian signal and a linear background. Using
incoherent Monte Carlo mixtures of the four subresonant modes we verified 
that our
procedure was able to correctly recover the generated mixtures of the four
modes.

The results for $K^-\pi^+\pi^+\pi^+\pi^-$ are
summarized and compared to the E687 results in Table~2. Taking into account the
correlation among the subresonant fractions, the calculated $\chi^2$ for the
hypothesis that the
results are consistent  with E687 is 6.5 (4 degrees of freedom). 
The four weighted histograms with fits are shown in
Fig.~2. Figure~2e is the weighted distribution for the sum of 
all subresonant modes. The goodness of fit is evaluated by calculating a 
$\chi^2$ for the hypothesis of consistency between the model predictions 
and observed data yields in each of the 8 submass bins. The calculated $\chi^2$
is 7.4 (4 degrees of freedom), 
with most of the $\chi^2$ contribution resulting from a poor Monte Carlo
simulation of the $\pi^+\pi^+$ spectrum for the $\overline{K^{*0}}\rho^0\pi^+$
mode. We assessed systematic errors by
individually varying the width of the submass bins corresponding 
to the $\rho$ and $\overline{K^{*0}}$ resonances by 20$\%$. The systematic
error is then estimated as the variance of the two measurements with varied
widths, along with the original measurement. Since our methods of calculating
subresonant fractions and inclusive branching ratios are distinct, statistical
and systematic errors are added in quadrature when normalizing our
subresonant fractions to other modes.
\begin{figure}[t]
\begin{center}
\includegraphics[width=4.5in]{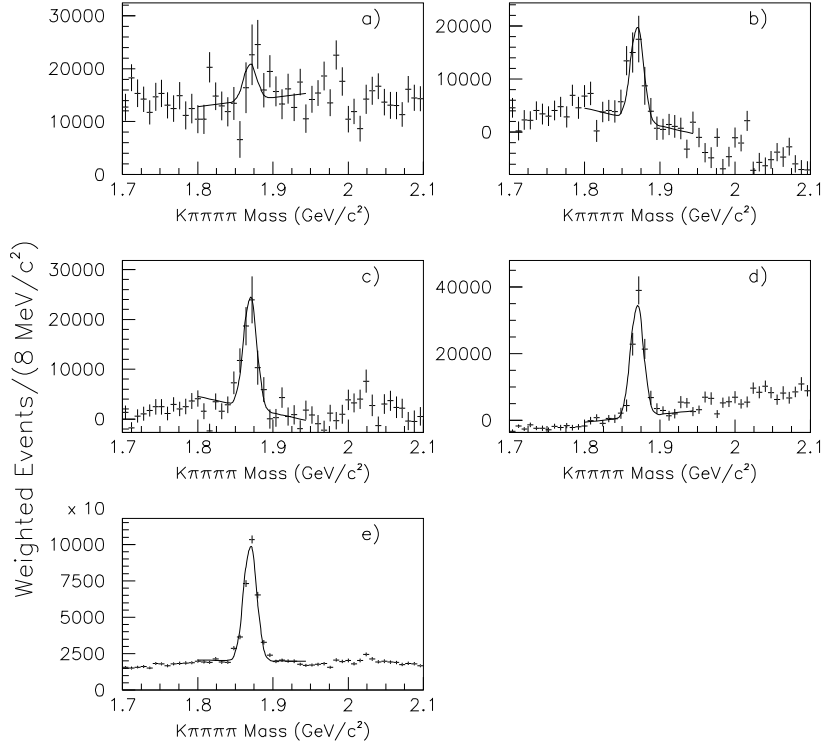}
\caption{$K^-\pi^+\pi^+\pi^+\pi^-$ weighted invariant mass for (a)
$(K^-\pi^+\pi^+\pi^+\pi^-)_{\textrm{NR}}$, (b) $\overline{K^{*0}}\pi^-\pi^+\pi^+$, 
(c) $K^-\rho^0\pi^+\pi^+$, (d) $\overline{K^{*0}}\rho^0\pi^+$,
(e) Inclusive sum of all four modes.}
\label{Fig 2}
\end{center}
\end{figure}
\begin{table}[h]
\begin{center}
\caption{Fractions relative to the inclusive
mode and comparison to previous measurements for the resonance substructure 
of the $D^+\!\rightarrow K^-\pi^+\pi^+\pi^+\pi^-$ decay mode. 
These values are not corrected for
unseen decay modes.}
\begin{tabular}{cccc} \hline \hline
Subresonant Mode& Fraction of K4$\pi$ & E687 Fraction [6] \\ 
\hline
$(K^-\pi^+\pi^+\pi^+\pi^-)_{\textrm{NR}}$&0.07$\pm$0.05$\pm$0.01& $<0.26(90\%$C.L.)\\
$\overline{K^{*0}}\pi^-\pi^+\pi^+$&0.21$\pm$0.04$\pm$0.06&0.42$\pm$0.14\\
$K^-\rho^0\pi^+\pi^+$&0.30$\pm$0.04$\pm$0.01&0.44$\pm$0.14\\
$\overline{K^{*0}}\rho^0\pi^+$&0.40$\pm$0.03$\pm$0.06&0.20$\pm$0.09\\
\hline
\hline
\end{tabular}
\end{center}
\end{table}

We follow a similar procedure for the $D_s^+\!\rightarrow K^+K^-\pi^+\pi^+\pi^-$,
treating the final state as an incoherent superposition of
the ($K^+K^-$) and ($\pi^+\pi^-$) resonances, as well as a nonresonant channel:
$\phi\pi^+\pi^+\pi^-$, $K^+K^-\rho\pi^+$, $\phi\rho\pi^+$ and 
$(K^+K^-\pi^+\pi^+\pi^-)_{\textrm{NR}}$. Each event is weighted by its value
in each of three submasses: ($K^+K^-$), ($\pi^+\pi^-$), and ($\pi^+\pi^+$), and
the weighted distributions are again fitted with a Gaussian signal and a linear
background. The results are summarized in Table~3 and 
are presented in Fig~3. The
goodness of fit is evaluated by calculating a $\chi^2$ for the hypothesis of
consistency between the model predictions and observed data yields in each of
the eight submass bins. 
The calculated $\chi^2$ is 10.2 (4 degrees of freedom), with most of the
$\chi^2$ contribution resulting from a poor Monte Carlo simulation of the $\pi^+\pi^+$ spectrum
in the nonresonant channel.
We assess systematic errors by
calculating the variance of our results with 20$\%$ variations in the width of
the submass bins corresponding to the $\rho$ and $\phi$
resonances.
\begin{figure}[t]
\begin{center}
\includegraphics[width=4.5in]{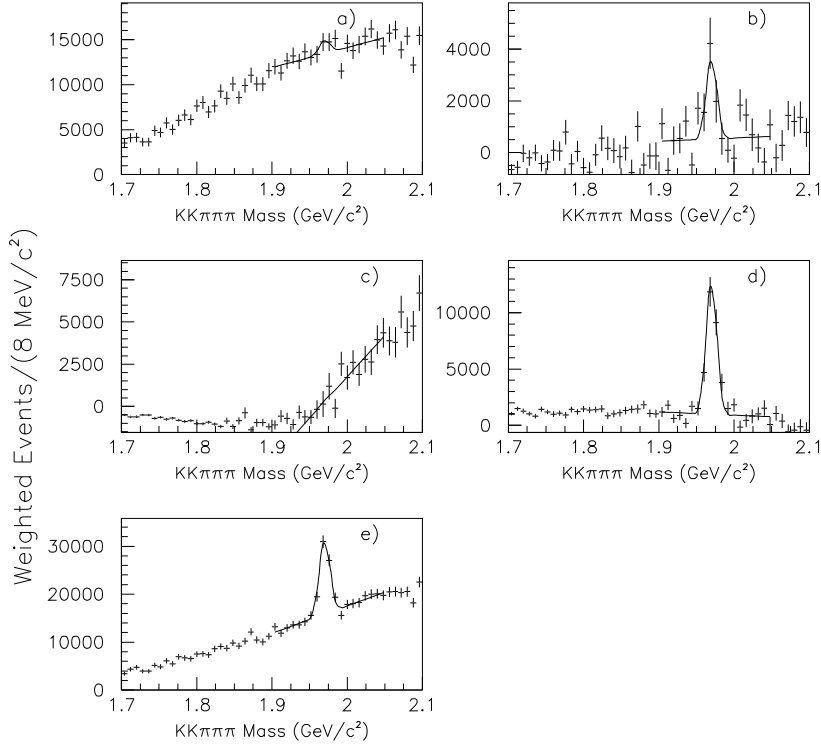}
\caption{$K^+K^-\pi^+\pi^+\pi^-$ weighted invariant mass for (a)
$(K^+K^-\pi^+\pi^+\pi^-)_{\textrm{NR}}$, (b) $\phi\pi^-\pi^+\pi^+$, 
(c) $K^+K^-\rho^0\pi^+$, (d) $\phi\rho^0\pi^+$,
(e) Inclusive sum of all four modes.}
\label{Fig 4}
\end{center}
\end{figure}
\begin{table}[b]
\begin{center}
\caption{Fractions relative to the inclusive
mode for the resonance substructure 
of the $D_s^+\!\rightarrow K^+K^-\pi^+\pi^+\pi^-$ decay mode. 
These values are not corrected for
unseen decay modes.}
\begin{tabular}{ccc} \hline \hline
Subresonant Mode & Fraction of 2K3$\pi$\\ 
\hline 
$(K^+K^-\pi^+\pi^+\pi^-)_{\textrm{NR}}$&0.10$\pm$0.06$\pm$0.05\\
$\phi\pi^-\pi^+\pi^+$&0.21$\pm$0.05$\pm$0.06\\
$K^+K^-\rho^0\pi^+$&$<$0.03 (90$\%$ C.L.)\\
$\phi\rho^0\pi^+$&0.75$\pm$0.06$\pm$0.04\\
\hline
\hline
\end{tabular}
\end{center}
\end{table}

In both resonant substructure analyses the dominant mode is of the form
vector-vector-pseudoscalar: $\overline{K^{*0}}\rho^0\pi^+$ and $\phi\rho^0\pi^+$
in the case of $K^-\pi^+\pi^+\pi^+\pi^-$ and $K^+K^-\pi^+\pi^+\pi^-$,
respectively. Given the phase space constraints for both of these decays, such a
result is unexpected. However, theoretical discussion of a vector-dominance
model for heavy flavor decays~[7] suggests that charm decays are dominated by
quasi-two-body decays in which the $W^{\pm}$ immediately
hadronizes into a charged pseudoscalar, vector, or axial vector meson. Thus
branching ratios of the form $D\!\rightarrow a_1(1260)^+X$ are of comparable value
to those observed for  $D\!\rightarrow \pi^+X$, when adjusted for phase space.
Such theoretical discussion raises the possibility that the resonant
substructure for both modes is dominated by a 
quasi-two-body decay involving the $a_1$: 
$\overline{K^{*0}}a_1^+$ and $\phi a_1^+$ for $K^-\pi^+\pi^+\pi^+\pi^-$ and 
$K^+K^-\pi^+\pi^+\pi^-$, respectively, where $a_1^+\!\rightarrow \rho^0\pi^+$.
Although the central value of the 
$a_1$ mass lies outside of phase space for both decays, these decay
modes are
allowed due to the large width of the $a_1$. However, 
the large width of the $a_1$ and its
position in phase space make the resonance difficult to detect directly.

To verify the subresonant decays are proceeding through $a_1$ we generate Monte
Carlo simulations of $D^+\!\rightarrow\overline{K^{*0}}a_1^+$ and
$D_s^+\!\rightarrow\phi a_1^+$, assuming the $a_1$  has a 
width of 400 MeV/$c^2$ and decays entirely as an S-wave,
and use our subresonant analysis
procedure explained above. In both cases the yield fractions 
in each of the subresonant modes from the 
$a_1$ Monte Carlo are similar to the reported fractions from the data,
with particular agreement in the case of $D_s^+\!\rightarrow\phi a_1^+$. Such
agreement suggests both channels may be dominated by a two-body
intermediate state involving the $a_1$.

Accepting the hypothesis that five-body modes are dominated by
quasi-two-body decays, we calculate branching ratios for the
decays $D^+\!\rightarrow\overline{K^{*0}}a_1^+$ and $D_s^+\!\rightarrow\phi a_1^+$
using the ratios of the observed fractions of $K^{*0}\rho\pi^+$ and $\phi\rho\pi$
from data (40$\%$ and 75$\%$) to the observed fractions from Monte Carlo
simulations of $D^+\!\rightarrow\overline{K^{*0}}a_1^+$ and
$D_s^+\!\rightarrow\phi a_1^+$ (70$\%$ and 78$\%$). Assuming the $a_1^+$ decays to $\rho^0\pi^+$
50$\%$ of the time and using the Particle Data Group $\phi$ and $K^{*0}$ branching
fractions~[11], the $D^+\!\rightarrow\overline{K^{*0}}a_1^+$ 
and $D_s^+\!\rightarrow\phi a_1^+$ branching fractions, including unseen decays,
are shown in Table~4. We assess systematic errors by increasing the width of
the
$a_1$  resonance in our generated Monte Carlo to 600 MeV/$c^2$, 
taking the systematic
error as the variance of our measurements with the two widths. 
\begin{table}[h]
\begin{center}
\caption{Inclusive branching ratios for $a_1^+$ states. These values are
corrected for unseen decay modes.}
\begin{tabular}{ccc} \hline \hline
Decay Mode & Fraction \\
\hline
$\frac{\Gamma(D^+\!\rightarrow \overline{K^{*0}}a_1^+)}
{\Gamma(D^+\!\rightarrow K^-\pi^+\pi^+)}$&
0.099$\pm$0.008$\pm$0.018\\
$\frac{\Gamma(D_s^+\!\rightarrow \phi a_1^+)}
{\Gamma(D_s^+\!\rightarrow K^+K-\pi^+)}$&
0.559$\pm$0.078$\pm$0.044\\
\hline
\hline
\end{tabular}
\end{center}
\end{table}

In conclusion we have measured the relative branching ratios of five-body and
three-body
charged hadronic decays of $D^+$ and $D_s^+$ and have presented the first
evidence of the decay mode $D^+\!\rightarrow K^+K^-\pi^+\pi^+\pi^-$. We have also
performed an analysis of the resonant substructure in the decays  
$D^+\!\rightarrow K^-\pi^+\pi^+\pi^+\pi^-$ and  
$D_s^+\!\rightarrow K^+K^-\pi^+\pi^+\pi^-$. Our analysis provides 
some evidence that both decays proceed through a quasi-two-body decay 
involving the
$a_1(1260)^+$ particle.

We acknowledge the assistance of the staffs of Fermi National
Accelerator Laboratory, the INFN of Italy, and the physics departments
of
the
collaborating institutions. This research was supported in part by the
U.~S.
National Science Foundation, the U.~S. Department of Energy, the Italian
Istituto Nazionale di Fisica Nucleare and
Ministero della Istruzione, Universit\`a e
Ricerca, the Brazilian Conselho Nacional de
Desenvolvimento Cient\'{\i}fico e Tecnol\'ogico, CONACyT-M\'exico, and
the Korea Research Foundation of
the Korean Ministry of Education.

\bibliographystyle{apsrev}
\bibliography{5body_plb}

\end{document}